\documentclass[aps,prl,twocolumn,groupedaddress,showpacs]{revtex4}

\usepackage{graphicx}
\usepackage{dcolumn}
\usepackage{bm}

\begin{document}


\title{An all-fiber source of pulsed twin beams at telecom band for quantum communication}



\author{Xueshi Guo$^1$, Xiaoying Li$^{1 *}$, Nannan Liu$^{\text{1}}$, Lei Yang$^{\text{1}}$, and Z. Y. Ou$^2$}

\affiliation{$^{\text{1}}$College of Precision Instrument and
Opto-electronics Engineering, Tianjin University, \\Key Laboratory of Optoelectronics Information Technology, Ministry of Education, Tianjin, 300072, China \\$^{\text{2}}$ Department of Physics, Indiana University-Purdue University Indianapolis, Indianapolis, IN 46202, USA}



\date{August 23, 2012}

\begin{abstract}
Motivated by the pursuit of a simple system to produce non-classical light sources for long-distance quantum communication, we generate for the first time an all-fiber source of pulsed twin beams in 1550 nm band by using a high gain fiber optical parametric amplifier. The noise of intensity difference of the twin beams is below the shot noise limit by 3.1 dB (10.4 dB after correction for losses). A detailed study reveals a number of limiting factors for higher noise reduction. Therefore, further noise reduction will be feasible once care is taken for these limiting factors.
\end{abstract}

\email[]{xiaoyingli@tju.edu.cn}

\maketitle


Robust and reliable quantum light sources are essential for large scale quantum information processing and quantum communication. The establishment of a quantum network requires long distance quantum communication with low loss. This restricts the operation wavelength at telecom band around 1550 nm. In quantum communication, information can be encoded discretely on each photon or continuously in the quadrature-phase amplitudes of an optical field~\cite{Tittel01,Braun05}. The former requires single-photon sources that can be generated with weak nonlinear interaction, and there has been a great amount of work in this area~\cite{lounis05}. The latter relies on strong quantum correlation between the amplitudes of optical fields for entanglement~\cite{Braun05,Reid09}. However, comparing with its discrete variable counterparts~\cite{Fasel2004,McMillan09,Yang11}, the efficient methods for generating continuous variable (CV) nonclassical light at telecom band need to be enriched~\cite{Mehmet11,Sharping01}.

The $\chi^{(2)}$ crystal based high gain optical parametric amplifier (OPA) is an efficient method for generating the
CV nonclassical light. Optical cavities are usually employed for the enhancement of nonlinear interaction in continuous wave (CW) operation \cite{wu86,ou92,Lau03,Vahlbruch08,Mehmet11}, which leads to not only complication in design and reduction in stability but also limited bandwidth. Another approach to enhance the nonlinearity is to use pulsed lasers as pump for OPA due to the available high peak power~\cite{Slusher87,Ayt90,Werner95,Wenger04,Zhang07}. Experimental systems based on the pulse pumped single-pass OPA are usually simpler than those with cavities but the measured levels of quantum-noise reduction fall short of those in CW cases, particularly for the case of ultrashort pump pulses~\cite{Lau03,Vahlbruch08,Ayt90,Zhang07} due to a lack of understanding about spectral properties of the systems~\cite{Wasi06}.

Recently, four-wave mixing (FWM) based on the $\chi^{(3)}$ nonlinearity in atoms or fibers was used to generate nonclassical light as well~\cite{Sharping01,Boy08}. The fiber systems, besides the wavelength match for telecom band and long interaction distance, has an obvious advantage over other approaches --- free from alignment. Hence, a fiber based system is more stable and integrable. Indeed, recent years have seen the growing interest in generating nonclassical light via the pulse pumped FWM in fibers~\cite{Li05a,Rarity05,Sharping01}. But so far, the FWM in fibers is in the low gain regime, where the noise reduction was less than 3 dB (even after correction for losses)~\cite{Sharping01}. For the high gain fiber optical parametric amplifiers (FOPA), the CW or quasi-CW pumped cases have been extensively studied~\cite{Hans02,Voss06,Mck12}, but the investigation of pulse pumped case is mostly on classical characteristics~\cite{Hans02}.

In this letter, using a high gain FOPA realized by pumping 300 m long dispersion shifted fiber (DSF) with a mode-locked fiber laser, we demonstrate a compact and robust all-fiber source of pulsed twin beams. The amplified signal and generated idler twin beams bear strong quantum correlation in intensity (photon-number) fluctuations, and the quantum noise level of their intensity difference is below the shot-noise limit (SNL) by 3.1 dB (10.4 dB after correction for losses). To the best of our knowledge, this is the highest quantum noise reduction obtained in a FOPA. Our source is not only capable of implementing quantum communication protocols in a quantum network within the current fiber network,  but also suitable for the applications
in high precision measurement, generation of continuous variable entanglement and quantum key distribution etc.~\cite{Braun05,Reid09,Gao98,Lau03,Su06,Funk02,Boy08}.

It should be noted that noise reduction via quasi-CW pumped near-degenerate FWM and via Kerr nonlinearity induced intensity dependent refractive index in fiber~\cite{Shelby86,Siberborn01,Hirosawa05} were achieved before but are limited mainly by the guided acoustic wave Brillouin scattering (GAWBS)~\cite{Shelby85}. In our experiment, however, the large detuning between signal (idler) and pump beams (about 2 THz), which is much greater than the frequency range of GAWBS, effectively eliminates the influence of GAWBS.

Our experimental setup is shown in Fig.1. The non-degenerate signal and idler twin beams are created in 300 m DSF via FWM by a strong pump pulse via $\chi^{(3)}$ nonlinear interaction with a seeded signal injection.
For the sake of convenience, we respectively label the beams having longer and shorter wavelengths as the signal and idler beams. A 90/10 fiber coupler is used to couple 90$\%$ of pump and 10$\%$ of signal pulses into the DSF. We create the pump and seeded signal pulses by
taking the 40 MHz train of 150-fs pulses centering
at 1560 nm from a mode-locked fiber laser, dispersing them with a grating
and then spectrally filtering them to obtain two synchronous
beams with a tunable wavelength separation
of about 15-20 nm. This arrangement gives 4 ps pump
pulses and 3 ps signal pulses, whose full width at half-maximum (FWHM) are about 0.9 and 1.2 nm, respectively.

\begin{figure}
\includegraphics[width=3in]{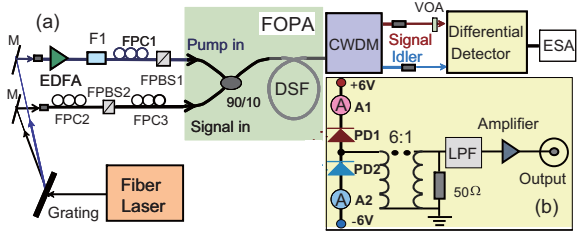}
\caption{\label{setup-v2}(Color online) (a) Experimental setup. M,
mirror; FPC, fiber polarization controller; FPBS, fiber polarization
beam splitter; VOA, variable optical attenuator; ESA, electronic spectrum analyzer. (b) The circuit of differential detector. PD, photodiodes; LPF, low pass filter; A, ampere-meters.
}
\end{figure}

To achieve the
required pump power, the pump pulses are amplified
by an erbium-doped fiber amplifier (EDFA), and further cleaned up spectrally with a
bandpass filter F1 having FWHM of about 0.9 or 0.5 nm. The polarization and power of the pump are controlled by a fiber polarization
controller (FPC1) and a fiber polarization beam
splitter (FPBS1). The central wavelength of pump is tuned to 1552.5 nm. Thus, for the DSF having zero dispersion wavelength (ZDW) of $\sim$1551 nm at the room temperature (300 K), the co-polarized FWM with a broad gain bandwidth is phase matched. For the input signal centering at 1569.8 nm, the intensity of signal seeded into DSF, $I_{in}$, is adjusted by using FPC2. The gain of FWM is maximized by matching the optical paths traversed by the
input pump and signal pulses and tuning
polarization of input signal with FPC3.

For an efficient collection of the generated signal and idler fields and their separation from the pump field, a four-channel coarse wavelength division multiplexer (CWDM) filter is placed after the DSF~\cite{guo12}. The central wavelengthes of the CWDM filter are 1511, 1531, 1551, and 1571 nm, respectively. For each channel, the transmission efficiency at the central wavelength is $\sim$80\%, the one-dB bandwidth is about 16 nm, which is 12 times broader than that of the input signal, and the isolation to the adjacent channels is $\sim40$ dB.

The signal and idler fields separated by the CWDM are directed to a differential detector system with the signal field passing through a variable optical attenuator (VOA) for the balance of the noise levels of the photocurrents from the signal and idler fields~\cite{Fabre89}. As shown in Fig. 1(b), the differential detection system is comprised of two photodiodes (OSI-InGaAs300), PD1 and PD2, with almost identical performances. The DC photocurrent of PD1 (PD2), $i_1$ ($i_{2}$), is monitored by an ampere-meter A1 (A2). The high frequency photocurrents coupled to a 50-$\Omega$ resistor with a 6:1 transformer is passed through a low-pass filter to reject the  repetition beat signal of the fiber laser, and then amplified by a low-noise electronic amplifier, whose output is fed to an electronic spectral analyzer (ESA).
The circuit allows one to measure the fluctuations of the difference of the photocurrents in ESA at a detection frequency of 3 MHz and a resolution bandwidth of 100 kHz. In our experiment, the maximized total detection efficiencies for signal and idler beams are about 55$\%$ and 58$\%$, respectively.

\begin{figure}
\includegraphics[width=3.25in]{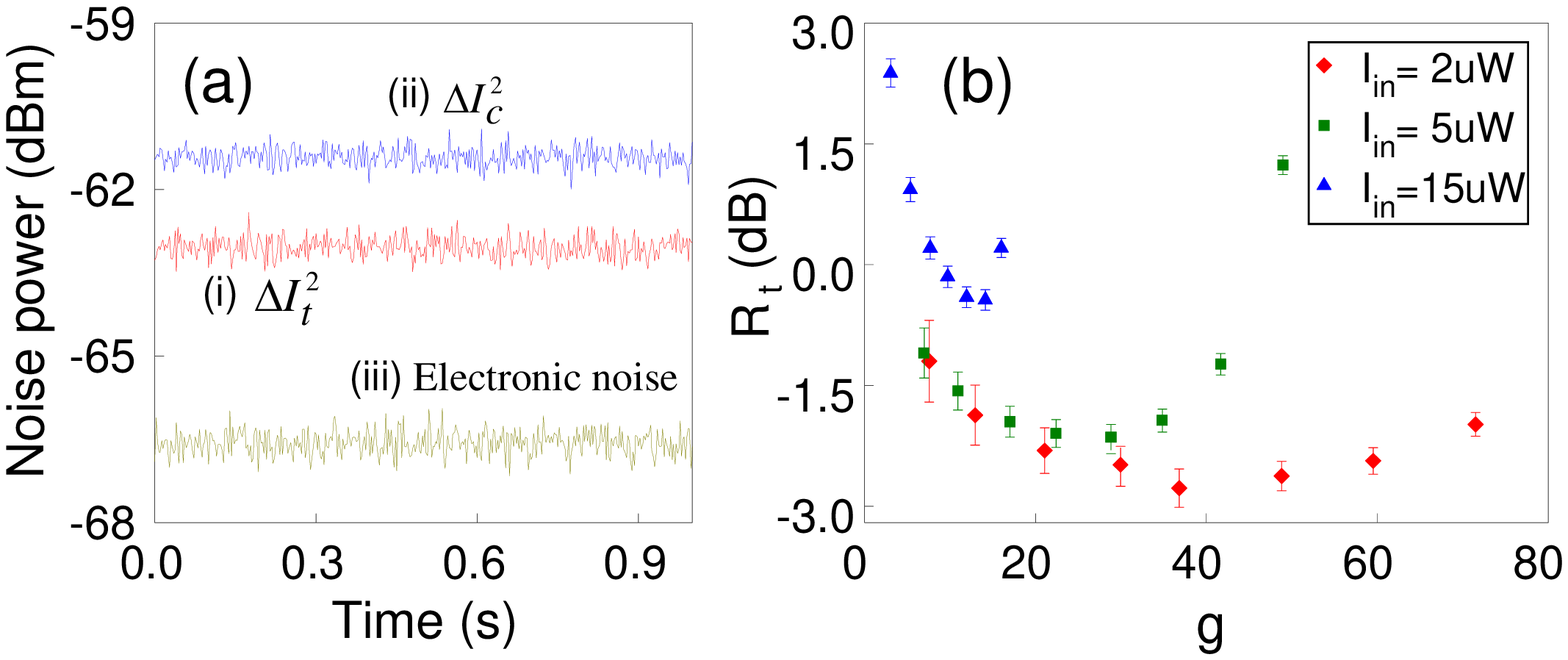}
\caption{\label{setup-v2}(Color online) (a) Noise level of the difference photocurrents for twin beams $\Delta I_t^2$ and corresponding SNL $\Delta I_c^2$. (b) The noise reduction $R_t\equiv \Delta I_t^2 /\Delta I_c^2$ versus the seeded parametric gain $g$ for different injection power $I_{in}$.
}
\end{figure}


Figure 2(a) shows the result of a typical measurement of the SNL $\Delta I_c^2$ and the noise level $\Delta I_t^2$ of difference photocurrents of the signal and idler fields when the injected signal intensity is $I_{in}= 2\mu W$, the average pump power is $1.7$ mW, and FWHM of pump is 0.9 nm. A directly measured quantum noise reduction of $R_t\equiv \Delta I_t^2/ \Delta I_c^2 \approx 1.7$ dB is obtained from Fig. 2(a). After correcting electronic noise of the instrumentation (trace (iii)), the result of $R_t$ is $2.6\pm0.2$ dB. The SNL is calibrated by illuminating the PDs with two beams derived from the fiber laser and using a VOA to match the DC current sum $i_1+i_2$ of the PDs with those for twin beams. We notice that the pump and input signal
have excess noise above SNL because of the noise from the fiber laser. So the SNL is obtained by using a balanced homodyne detection technique to eliminate the excess noise~\cite{Yuen83}.

In order to investigate the performance of the system and the limiting factors for quantum noise reduction, we first measure $R_t$ of twin beams versus the seeded parametric gain
$g \equiv I_s/I_{in}$ for $I_{in}$ = 2, 5, 15 $\mu$W, respectively, as shown in Fig. 2(b). $I_s$ and $I_{in}$ are the output and input signal intensities of our FOPA. The gain $g$ is adjusted by varying the pump power. Each data point is obtained by slightly varying the efficiency of the signal beam via VOA to maximize the observed noise reduction and by subtracting the electronic noise of the instrumentation. Figure 2(b) clearly shows that $R_t$ falls below the SNL for some ranges of gain and injected power, indicating the existence of quantum noise correlation between the signal and idler beams. For each set of data, the changing tendency of $R_t$ is similar: with the increase of $g$, the graph presents a descending curve in the beginning, then starts to ascend when $R_t$ reaches the minimum at a certain turning point of $g$. With the decrease of $I_{in}$, the minimum $R_t$ decreases, and the corresponding turning point of $g$ increases. So there are some detrimental effects at high injection and high pump power that limit the amount of noise reduction.

First of all, the reason for the increase of $R_t$ with the injected power is straightforward: as we mentioned earlier, the injected signal beam originated from fiber laser is not shot-noise limited. Such excess noise increases with the power and lifts up the level of $R_t$. Notice that for the injection power of 15 $\mu W$, whose intensity noise is $\sim12$ dB above SNL, the measured $R_t$ is above SNL in the low gain regime ($g<7$), because the quantum correlation between the signal and idler beams created at low gain FWM is not strong enough to bring the subtracted noise down below SNL. This is different from the result in Ref.~\cite{Sharping01}, in which the laser system was shot noise limited and $R_t$ was below SNL for $g<3$.

\begin{figure}
\includegraphics[width=3in]{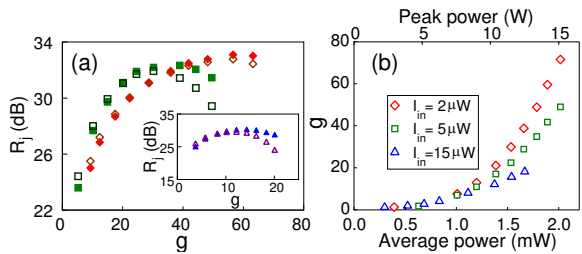}
\caption{(Color online) (a) The noise level of individual signal and idler beams $R_{j}$ ($j=s,i$) relative to SNL versus $g$ for $I_{in}=2$ (diamonds), $I_{in}=5$ (squares), and $I_{in}=15$ $\mu$W (triangles). (b) The parametric gain $g$ versus the pump power.
}
\end{figure}

To understand the rise of $R_t$ at high gain, we measure the noise levels $R_s, R_i$ in reference to SNL of individual signal and idler beams as a function of the gain when the detection efficiencies of the two beams are adjusted to be equal, as shown in Fig. 3(a) where the solid filled data points are for signal beam and the hollow data points are for idler beam. Comparing Figs. 2(b) and 3(a), we find that when $g$ is in the vicinity of the turning points in Fig. 2(b), within which the minimum $R_t$ is observable, $R_s$ and $R_i$ in Fig. 3(a) are about the same; when $g$ is less than the turning point, the values of both $R_s$ and $R_i$ increase with $g$, and the noise of the generated idler beam $R_i$ (hollow points) is higher than that of the amplified signal beam $R_s$ (solid points), which agree with the theoretical result obtained by neglecting the pump depletion~\cite{guo12}. However, when $g$ is greater than the turning point, the values of both $R_s$ and $R_i$ decrease with $g$, particularly, for $I_{in}$ at a higher power. We think this is caused by the gain saturation~\cite{Inoue02}. Therefore, the results in Figs. 2(b) and 3(a) indicate gain saturation will degrade the correlation between signal and idler beams, which prevent $R_t$ from further reduction.

To confirm the argument above more directly, we look at the classical parametric gain of the FOPA as a function of the pump power for different injected signal powers in Fig. 3(b). If there were no saturation, $g$ would be independent of $I_{in}$, but Fig. 3(b) shows that for a certain pump power, $g$ decreases with the increase of $I_{in}$, particularly in the high gain regime. We next characterize the temporal mode of twin beams by measuring their pulse duration and time-bandwidth product (TBP). The results show the pulse duration of both the signal and idler beams decreases with the increase of $g$; the TBP of signal, about transform limited, does not vary with $g$, but the TBP of idler beam increases with $g$ because the chirps induced by dispersion of DSF and cross phase modulation of pump in idler band can not cancel out~\cite{Agrawal}. Therefore, we think the reasons responsible for the gain saturation are: (i) the depletion of pump power, and (ii) the decrease of temporal overlap between the pump and twin beams due to the increased TBP of idler beam and group velocity walk-off.

\begin{figure}
\includegraphics[width=3in]{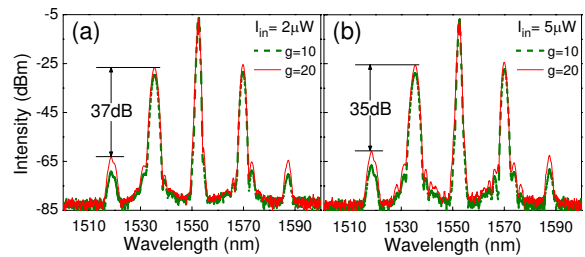}
\caption{(Color online) Spectra of the output of FOPA for (a) $I_{in}=2$ and (b) $I_{in}=5$ $\mu$W, respectively.
}
\end{figure}

Since gain saturation can be mitigated by lengthening the pulse duration to improve the temporal mode matching, we repeat the above measurements by passing the pump through a filter (F1) with an FWHM of about 0.5 nm. In this case, the difference in the gains for $I_{in}$ with different intensity obtained under a certain pump power is less obvious than that in Fig. 3(b). Accordingly, the measurement of quantum noise fluctuation shows that for a certain $I_{in}$, the turning point of $g$ is greater than that in Figs. 2(b) and 3(a) due to the mitigatory gain saturation. However, in contrast to the theoretical prediction~\cite{guo12}, $R_t$ measured for a certain $I_{in}$ and $g$ is always higher than that in Fig. 2(b).

To figure out why the measured noise reduction $R_t$ becomes worse for pump with a narrower FWHM, we then analyze the intensity of high-order FWM by taking the CWDM out and directly sending the output of DSF into an optical spectrum analyzer. As shown in Fig. 4, the intensity of high-order FWM, which increases with $g$ and $I_{in}$, is more than 30 dB less than that the first order FWM. However, comparing with the results obtain for pump with FWHM of 0.9 nm, the intensity of high-order FWM is slightly enhanced. For instance, for the case of $g=20$ obtained by using the pump with FWHM of 0.9 nm and signal with $I_{in}=2$ $\mu$W, the ratio between the intensity of second-order and first-order FWMs is about 2 dB less than that in Fig. 4(a). Therefore, we think the increased high-order FWM will prevent $R_t$ from further decreasing, although its intensity is still very weak.

\begin{figure}
\includegraphics[width=3in]{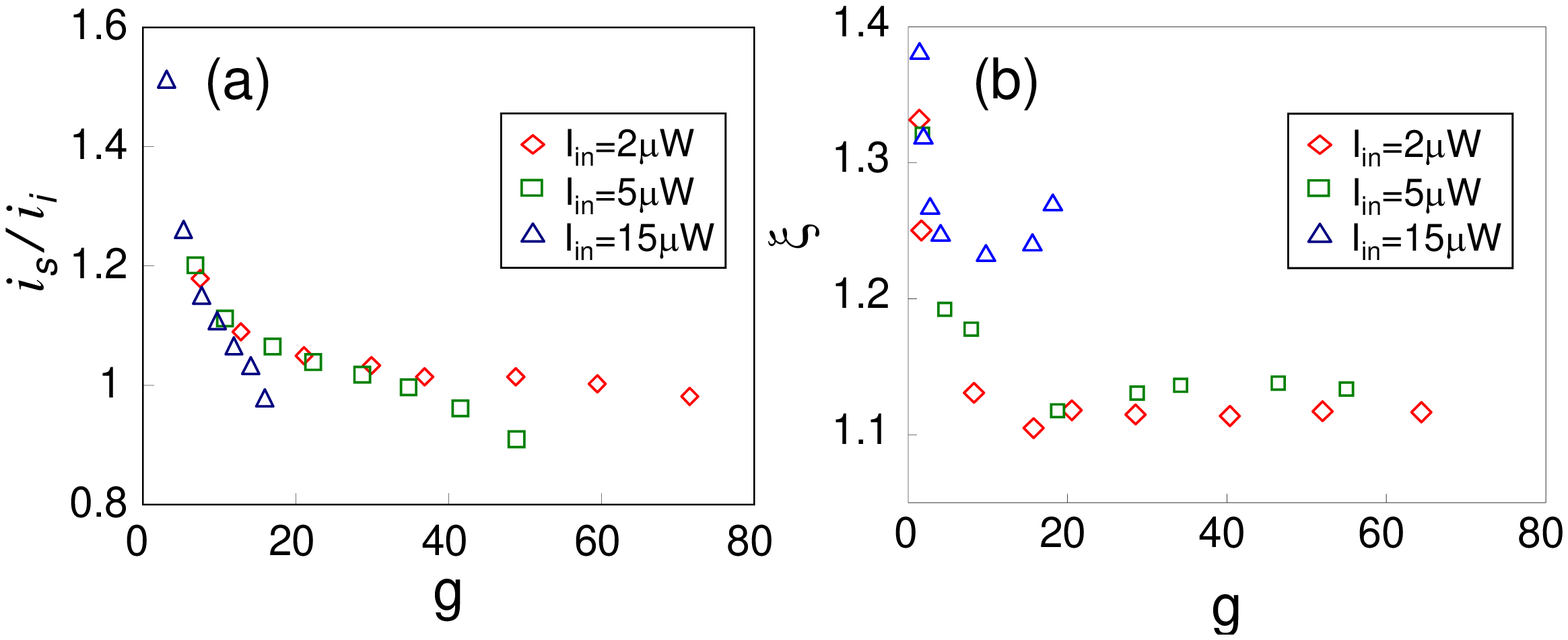}
\caption{(Color online) (a) The ratio of the DC currents $i_1/i_2$ versus $g$. (b) The ratio $\xi= (I_s-I_{in})/{I_i}$ versus $g$.
}
\end{figure}

In order to accurately evaluate the detection efficiencies of signal and idler beams, during data taking for Fig. 2(b), the DC currents of the two PDs, $i_1$ and $i_2$, are also recorded, and the ratio $i_1/i_2$ versus $g$ is shown in Fig. 5(a). Comparing Figs. 2(a) and 5(a), we find the minimum $R_t$ is observed under the condition of $i_1/i_2\approx1.01$, which is very close to ideal condition $i_1/i_2=1$ for eliminating side effect induced by the excess noise of input signal~\cite{guo12}. Moreover, it is worth noting that for an ideal FOPA, we should have $\frac{I_s-I_{in}}{I_i}=1$ (the unit of $I_{s(i)}$ and $I_{in}$ is photon number), where $I_i$ denotes the intensity of idler beam, and $I_{s(i)}$ are corrected by the detection efficiencies of signal and idler beams. However, we find the relation $(I_s-I_{in})/{I_i}=1$ is not satisfied in our experiment. To illustrate this phenomena clearly, we plot the ratio $\xi= (I_s-I_{in})/{I_i}$ versus $g$ in Fig. 5(b), showing the photon number asymmetry between the amplified signal and idler beams. Obviously, we have $\xi>1$ for all the gain values. Moreover, in the low gain regime, the changing trend for each set of data is similar: the ratio $\xi$ decreases with $g$; in the high gain regime, the variation of $\xi$ is not evident, however, the value of $\xi$ will start to increase when the gain saturation becomes distinct. We believe the observed asymmetry of photon number is caused by the parasitical Raman effect and gain saturation~\cite{Hsieh07,Peucheret12}.

\begin{figure}
\includegraphics[width=3in]{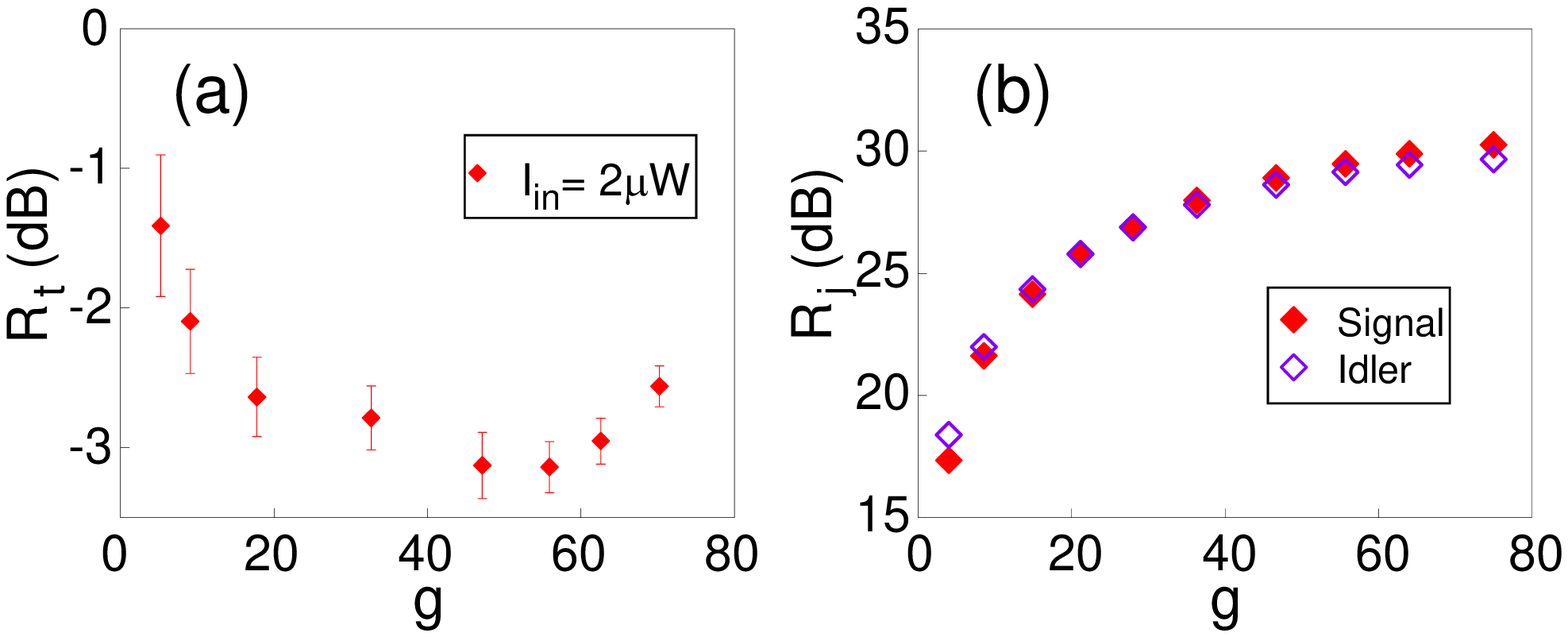}
\caption{\label{setup-v2}(Color online) (a) $R_t$ and (b) $R_{j}$ ($j=s,i$) versus $g$ for injection power $I_{in}=2$ $\mu$W with the DSF at 77 K.
}
\end{figure}

To mitigate the influence of Raman effect, we measure $R_t$ versus $g$ by immersing the 300 m DSF in liquid nitrogen (77 K). In this case, the cooling induced loss is less than 2$\%$, and the ZDW of DSF is about 3 nm shorter than that at 300 K. To ensure the gain bandwidth of FWM is about the same as that at 300 K, we accordingly adjust the central wavelengths of pump to 1549 nm. During the measurement, the FWHM of pump is 0.9 nm, $I_{in}$ is 2 $\mu$W, and the detuning between signal and pump beam does not change. We find that the classical gain character of the FOPA is the same as that at 300 K, but the quantum noise fluctuation of twin beams changes. As shown in Fig. 6(a), for a certain $g$, $R_t$ obtained at 77 K is smaller than that in Fig. 2(b). When $g$ is about 56, $R_t$ falls below the SNL by $3.1$ dB. If we correct the measured $R_t$ by the detection efficiencies, which are 53$\%$ and 57$\%$ for signal and idler beams, respectively, $R_t$ would be 10.4 dB. Moreover, from Fig. 6(b), we find that for a certain $g$, the measured $R_j$ ($j=s,i$) of individual beam is less than that in Fig. 3(a) because the noise induced by Raman scattering is suppressed~\cite{Voss06}.

Our experimental investigation shows that the fundamental factors limiting the noise reduction of twin beams are: (i) the noise of seeded signal,(ii) gain saturation, and (iii) parasitical nonlinear effects, including high-order FWM and Raman effect. The first and second factors can be respectively suppressed by suppressing the noise of laser system and lowering the power of $I_{in}$, while the third factor can be avoided by engineering the phase matching of FOPA so that the gain bandwidth of FWM is narrow and the detuning of pump-signal is large enough~\cite{Rarity05,Palmett07}. On the other hand, the practical issues affecting the measured $R_t$ are the collection and detection efficiencies. Apart from using CWDM filters and PDs with better performance, the observed amount and bandwidth of noise reduction can be further improved by optimizing the electrical gain of PDs and by increasing the response bandwidth detection system.

In conclusion, we have developed a compact and integrable all-fibre source of pulsed twin beams. When the seeded parametric gain is about 56, the noise of intensity difference of the twin beams drops below the SNL by 3.1 dB (10.4 dB when corrected for losses). Thorough analysis reveals that even higher noise reduction will be feasible. Our investigation shows that a pulse pumped high gain FOPA provides a new means of generating the CV nonclassical light with a simple system.

\begin{acknowledgments}
We are grateful to Prof. P. Kumar and Prof. P. Lam for useful discussion. This work was supported in part by the State Key Development Program for Basic Research of China (No. 2010CB923101).
\end{acknowledgments}

\bibliographystyle{apsrev}


\begin{thebibliography}{40}
\expandafter\ifx\csname natexlab\endcsname\relax\def\natexlab#1{#1}\fi
\expandafter\ifx\csname bibnamefont\endcsname\relax
  \def\bibnamefont#1{#1}\fi
\expandafter\ifx\csname bibfnamefont\endcsname\relax
  \def\bibfnamefont#1{#1}\fi
\expandafter\ifx\csname citenamefont\endcsname\relax
  \def\citenamefont#1{#1}\fi
\expandafter\ifx\csname url\endcsname\relax
  \def\url#1{\texttt{#1}}\fi
\expandafter\ifx\csname urlprefix\endcsname\relax\def\urlprefix{URL }\fi
\providecommand{\bibinfo}[2]{#2}
\providecommand{\eprint}[2][]{\url{#2}}

\bibitem[{\citenamefont{Tittel and Weihs}(2001)}]{Tittel01}
\bibinfo{author}{\bibfnamefont{W.}~\bibnamefont{Tittel}} \bibnamefont{and}
  \bibinfo{author}{\bibfnamefont{G.}~\bibnamefont{Weihs}},
  \bibinfo{journal}{Quantum Information and Computation}
  \textbf{\bibinfo{volume}{1}}, \bibinfo{pages}{3} (\bibinfo{year}{2001}).

\bibitem[{\citenamefont{Braunstein and Loock}(2005)}]{Braun05}
\bibinfo{author}{\bibfnamefont{S.~L.} \bibnamefont{Braunstein}}
  \bibnamefont{and} \bibinfo{author}{\bibfnamefont{P.~V.} \bibnamefont{Loock}},
  \bibinfo{journal}{Rev. Modern Phys.} \textbf{\bibinfo{volume}{77}},
  \bibinfo{pages}{513} (\bibinfo{year}{2005}).

\bibitem[{\citenamefont{Lounis and Orrit}(2005)}]{lounis05}
\bibinfo{author}{\bibfnamefont{B.}~\bibnamefont{Lounis}} \bibnamefont{and}
  \bibinfo{author}{\bibfnamefont{M.}~\bibnamefont{Orrit}},
  \bibinfo{journal}{Rep. Prog. Phys} \textbf{\bibinfo{volume}{68}},
  \bibinfo{pages}{1129} (\bibinfo{year}{2005}).

\bibitem[{\citenamefont{Reid et~al.}(2009)\citenamefont{Reid, Drummond, Bowen,
  Cavalcanti, Lam, Bachor, Andersen, and Leuchs}}]{Reid09}
\bibinfo{author}{\bibfnamefont{M.~D.} \bibnamefont{Reid}},
  \bibinfo{author}{\bibfnamefont{P.~D.} \bibnamefont{Drummond}},
  \bibinfo{author}{\bibfnamefont{W.~P.} \bibnamefont{Bowen}},
  \bibinfo{author}{\bibfnamefont{E.~G.} \bibnamefont{Cavalcanti}},
  \bibinfo{author}{\bibfnamefont{P.~K.} \bibnamefont{Lam}},
  \bibinfo{author}{\bibfnamefont{H.~A.} \bibnamefont{Bachor}},
  \bibinfo{author}{\bibfnamefont{U.~L.} \bibnamefont{Andersen}},
  \bibnamefont{and} \bibinfo{author}{\bibfnamefont{G.}~\bibnamefont{Leuchs}},
  \bibinfo{journal}{Rev. Modern Phys.} \textbf{\bibinfo{volume}{81}},
  \bibinfo{pages}{1727} (\bibinfo{year}{2009}).

\bibitem[{\citenamefont{Fasel et~al.}(2004)\citenamefont{Fasel, Alibart,
  Tanzilli, Baldi, Beveratos, Gisin, and Zbinden}}]{Fasel2004}
\bibinfo{author}{\bibfnamefont{S.}~\bibnamefont{Fasel}},
  \bibinfo{author}{\bibfnamefont{O.}~\bibnamefont{Alibart}},
  \bibinfo{author}{\bibfnamefont{S.}~\bibnamefont{Tanzilli}},
  \bibinfo{author}{\bibfnamefont{P.}~\bibnamefont{Baldi}},
  \bibinfo{author}{\bibfnamefont{A.}~\bibnamefont{Beveratos}},
  \bibinfo{author}{\bibfnamefont{N.}~\bibnamefont{Gisin}}, \bibnamefont{and}
  \bibinfo{author}{\bibfnamefont{H.}~\bibnamefont{Zbinden}},
  \bibinfo{journal}{New J. Phys.} \textbf{\bibinfo{volume}{6}},
  \bibinfo{pages}{163} (\bibinfo{year}{2004}).

\bibitem[{\citenamefont{McMillan et~al.}(2009)\citenamefont{McMillan, Fulconis,
  Halder, Xiong, Rarity, and Wadsworth}}]{McMillan09}
\bibinfo{author}{\bibfnamefont{A.}~\bibnamefont{McMillan}},
  \bibinfo{author}{\bibfnamefont{J.}~\bibnamefont{Fulconis}},
  \bibinfo{author}{\bibfnamefont{M.}~\bibnamefont{Halder}},
  \bibinfo{author}{\bibfnamefont{C.}~\bibnamefont{Xiong}},
  \bibinfo{author}{\bibfnamefont{J.}~\bibnamefont{Rarity}}, \bibnamefont{and}
  \bibinfo{author}{\bibfnamefont{W.}~\bibnamefont{Wadsworth}},
  \bibinfo{journal}{Opt. Express} \textbf{\bibinfo{volume}{17}},
  \bibinfo{pages}{6156} (\bibinfo{year}{2009}).

\bibitem[{\citenamefont{Yang et~al.}(2011)\citenamefont{Yang, Ma, Guo, Cui, and
  Li}}]{Yang11}
\bibinfo{author}{\bibfnamefont{L.}~\bibnamefont{Yang}},
  \bibinfo{author}{\bibfnamefont{X.}~\bibnamefont{Ma}},
  \bibinfo{author}{\bibfnamefont{X.}~\bibnamefont{Guo}},
  \bibinfo{author}{\bibfnamefont{L.}~\bibnamefont{Cui}}, \bibnamefont{and}
  \bibinfo{author}{\bibfnamefont{X.}~\bibnamefont{Li}}, \bibinfo{journal}{Phys.
  Rev. A} \textbf{\bibinfo{volume}{83}}, \bibinfo{pages}{053843}
  (\bibinfo{year}{2011}).

\bibitem[{\citenamefont{Mehmet et~al.}(2011)\citenamefont{Mehmet, Ast, Eberle,
  Steinlechner, Vahlbruch, and Schnabel}}]{Mehmet11}
\bibinfo{author}{\bibfnamefont{M.}~\bibnamefont{Mehmet}},
  \bibinfo{author}{\bibfnamefont{S.}~\bibnamefont{Ast}},
  \bibinfo{author}{\bibfnamefont{T.}~\bibnamefont{Eberle}},
  \bibinfo{author}{\bibfnamefont{S.}~\bibnamefont{Steinlechner}},
  \bibinfo{author}{\bibfnamefont{H.}~\bibnamefont{Vahlbruch}},
  \bibnamefont{and} \bibinfo{author}{\bibfnamefont{R.}~\bibnamefont{Schnabel}},
  \bibinfo{journal}{Opt. Express} \textbf{\bibinfo{volume}{19}},
  \bibinfo{pages}{25763} (\bibinfo{year}{2011}).

\bibitem[{\citenamefont{Sharping et~al.}(2001)\citenamefont{Sharping,
  Fiorentino, and Kumar}}]{Sharping01}
\bibinfo{author}{\bibfnamefont{J.~E.} \bibnamefont{Sharping}},
  \bibinfo{author}{\bibfnamefont{M.}~\bibnamefont{Fiorentino}},
  \bibnamefont{and} \bibinfo{author}{\bibfnamefont{P.}~\bibnamefont{Kumar}},
  \bibinfo{journal}{Opt. Lett.} \textbf{\bibinfo{volume}{26}},
  \bibinfo{pages}{367} (\bibinfo{year}{2001}).

\bibitem[{\citenamefont{Wu et~al.}(1986)\citenamefont{Wu, Kimble, Hall, and
  Wu}}]{wu86}
\bibinfo{author}{\bibfnamefont{L.~A.} \bibnamefont{Wu}},
  \bibinfo{author}{\bibfnamefont{H.~J.} \bibnamefont{Kimble}},
  \bibinfo{author}{\bibfnamefont{J.~L.} \bibnamefont{Hall}}, \bibnamefont{and}
  \bibinfo{author}{\bibfnamefont{H.}~\bibnamefont{Wu}}, \bibinfo{journal}{Phys.
  Rev. Lett.} \textbf{\bibinfo{volume}{57}}, \bibinfo{pages}{2520}
  (\bibinfo{year}{1986}).

\bibitem[{\citenamefont{Ou et~al.}(1992)\citenamefont{Ou, Pereira, Kimble, and
  Peng}}]{ou92}
\bibinfo{author}{\bibfnamefont{Z.~Y.} \bibnamefont{Ou}},
  \bibinfo{author}{\bibfnamefont{S.~F.} \bibnamefont{Pereira}},
  \bibinfo{author}{\bibfnamefont{H.~J.} \bibnamefont{Kimble}},
  \bibnamefont{and} \bibinfo{author}{\bibfnamefont{K.~C.} \bibnamefont{Peng}},
  \bibinfo{journal}{Phys. Rev. Lett.} \textbf{\bibinfo{volume}{68}},
  \bibinfo{pages}{3663} (\bibinfo{year}{1992}).

\bibitem[{\citenamefont{Laurat et~al.}(2003)\citenamefont{Laurat, Coudreau,
  Treps, Ma\^{\i}tre, and Fabre}}]{Lau03}
\bibinfo{author}{\bibfnamefont{J.}~\bibnamefont{Laurat}},
  \bibinfo{author}{\bibfnamefont{T.}~\bibnamefont{Coudreau}},
  \bibinfo{author}{\bibfnamefont{N.}~\bibnamefont{Treps}},
  \bibinfo{author}{\bibfnamefont{A.}~\bibnamefont{Ma\^{\i}tre}},
  \bibnamefont{and} \bibinfo{author}{\bibfnamefont{C.}~\bibnamefont{Fabre}},
  \bibinfo{journal}{Phys. Rev. Lett.} \textbf{\bibinfo{volume}{91}},
  \bibinfo{pages}{213601} (\bibinfo{year}{2003}).

\bibitem[{\citenamefont{Vahlbruch et~al.}(2008)\citenamefont{Vahlbruch, Mehmet,
  Chelkowski, Hage, Franzen, Lastzka, Go\ss{}ler, Danzmann, and
  Schnabel}}]{Vahlbruch08}
\bibinfo{author}{\bibfnamefont{H.}~\bibnamefont{Vahlbruch}},
  \bibinfo{author}{\bibfnamefont{M.}~\bibnamefont{Mehmet}},
  \bibinfo{author}{\bibfnamefont{S.}~\bibnamefont{Chelkowski}},
  \bibinfo{author}{\bibfnamefont{B.}~\bibnamefont{Hage}},
  \bibinfo{author}{\bibfnamefont{A.}~\bibnamefont{Franzen}},
  \bibinfo{author}{\bibfnamefont{N.}~\bibnamefont{Lastzka}},
  \bibinfo{author}{\bibfnamefont{S.}~\bibnamefont{Go\ss{}ler}},
  \bibinfo{author}{\bibfnamefont{K.}~\bibnamefont{Danzmann}}, \bibnamefont{and}
  \bibinfo{author}{\bibfnamefont{R.}~\bibnamefont{Schnabel}},
  \bibinfo{journal}{Phys. Rev. Lett.} \textbf{\bibinfo{volume}{100}},
  \bibinfo{pages}{033602} (\bibinfo{year}{2008}).

\bibitem[{\citenamefont{Slusher et~al.}(1987)\citenamefont{Slusher, Grangier,
  LaPorta, Yurke, and Potasek}}]{Slusher87}
\bibinfo{author}{\bibfnamefont{R.~E.} \bibnamefont{Slusher}},
  \bibinfo{author}{\bibfnamefont{P.}~\bibnamefont{Grangier}},
  \bibinfo{author}{\bibfnamefont{A.}~\bibnamefont{LaPorta}},
  \bibinfo{author}{\bibfnamefont{B.}~\bibnamefont{Yurke}}, \bibnamefont{and}
  \bibinfo{author}{\bibfnamefont{M.~J.} \bibnamefont{Potasek}},
  \bibinfo{journal}{Phys. Rev. Lett.} \textbf{\bibinfo{volume}{59}},
  \bibinfo{pages}{2566} (\bibinfo{year}{1987}).

\bibitem[{\citenamefont{Ayt\"{u}r and Kumar}(1990)}]{Ayt90}
\bibinfo{author}{\bibfnamefont{O.}~\bibnamefont{Ayt\"{u}r}} \bibnamefont{and}
  \bibinfo{author}{\bibfnamefont{P.}~\bibnamefont{Kumar}},
  \bibinfo{journal}{Phys. Rev. Lett.} \textbf{\bibinfo{volume}{65}},
  \bibinfo{pages}{1551} (\bibinfo{year}{1990}).

\bibitem[{\citenamefont{Werner et~al.}(1995)\citenamefont{Werner, Raymer, Beck,
  and Drummond}}]{Werner95}
\bibinfo{author}{\bibfnamefont{M.~J.} \bibnamefont{Werner}},
  \bibinfo{author}{\bibfnamefont{M.~G.} \bibnamefont{Raymer}},
  \bibinfo{author}{\bibfnamefont{M.}~\bibnamefont{Beck}}, \bibnamefont{and}
  \bibinfo{author}{\bibfnamefont{P.~D.} \bibnamefont{Drummond}},
  \bibinfo{journal}{Phys. Rev. A} \textbf{\bibinfo{volume}{52}},
  \bibinfo{pages}{4202} (\bibinfo{year}{1995}).

\bibitem[{\citenamefont{Wenger et~al.}(2004)\citenamefont{Wenger,
  Tualle-Brouri, and Grangier}}]{Wenger04}
\bibinfo{author}{\bibfnamefont{J.}~\bibnamefont{Wenger}},
  \bibinfo{author}{\bibfnamefont{R.}~\bibnamefont{Tualle-Brouri}},
  \bibnamefont{and} \bibinfo{author}{\bibfnamefont{P.}~\bibnamefont{Grangier}},
  \bibinfo{journal}{Opt. Lett.} \textbf{\bibinfo{volume}{29}},
  \bibinfo{pages}{1267} (\bibinfo{year}{2004}).

\bibitem[{\citenamefont{Zhang et~al.}(2007)\citenamefont{Zhang, Furuta, Okubo,
  Takahashi, and Hirano}}]{Zhang07}
\bibinfo{author}{\bibfnamefont{Y.}~\bibnamefont{Zhang}},
  \bibinfo{author}{\bibfnamefont{T.}~\bibnamefont{Furuta}},
  \bibinfo{author}{\bibfnamefont{R.}~\bibnamefont{Okubo}},
  \bibinfo{author}{\bibfnamefont{K.}~\bibnamefont{Takahashi}},
  \bibnamefont{and} \bibinfo{author}{\bibfnamefont{T.}~\bibnamefont{Hirano}},
  \bibinfo{journal}{Phys. Rev. A} \textbf{\bibinfo{volume}{76}},
  \bibinfo{pages}{012304} (\bibinfo{year}{2007}).

\bibitem[{\citenamefont{Wasilewski et~al.}(2006)\citenamefont{Wasilewski,
  Lvovsky, Banaszek, and Radzewicz}}]{Wasi06}
\bibinfo{author}{\bibfnamefont{W.}~\bibnamefont{Wasilewski}},
  \bibinfo{author}{\bibfnamefont{A.~I.} \bibnamefont{Lvovsky}},
  \bibinfo{author}{\bibfnamefont{K.}~\bibnamefont{Banaszek}}, \bibnamefont{and}
  \bibinfo{author}{\bibfnamefont{C.}~\bibnamefont{Radzewicz}},
  \bibinfo{journal}{Phys. Rev. A} \textbf{\bibinfo{volume}{73}},
  \bibinfo{pages}{063819} (\bibinfo{year}{2006}).

\bibitem[{\citenamefont{Boyer et~al.}(2008)\citenamefont{Boyer, Marino, Pooser,
  and Lett£¬}}]{Boy08}
\bibinfo{author}{\bibfnamefont{V.}~\bibnamefont{Boyer}},
  \bibinfo{author}{\bibfnamefont{A.~M.} \bibnamefont{Marino}},
  \bibinfo{author}{\bibfnamefont{R.~C.} \bibnamefont{Pooser}},
  \bibnamefont{and} \bibinfo{author}{\bibfnamefont{P.~D.}
  \bibnamefont{Lett£¬}}, \bibinfo{journal}{Science}
  \textbf{\bibinfo{volume}{321}}, \bibinfo{pages}{544} (\bibinfo{year}{2008}).

\bibitem[{\citenamefont{Li et~al.}(2005)\citenamefont{Li, Voss, Sharping, and
  Kumar}}]{Li05a}
\bibinfo{author}{\bibfnamefont{X.}~\bibnamefont{Li}},
  \bibinfo{author}{\bibfnamefont{P.~L.} \bibnamefont{Voss}},
  \bibinfo{author}{\bibfnamefont{J.~E.} \bibnamefont{Sharping}},
  \bibnamefont{and} \bibinfo{author}{\bibfnamefont{P.}~\bibnamefont{Kumar}},
  \bibinfo{journal}{Phys. Rev. Lett.} \textbf{\bibinfo{volume}{94}},
  \bibinfo{pages}{053601} (\bibinfo{year}{2005}).

\bibitem[{\citenamefont{Rarity et~al.}(2005)\citenamefont{Rarity, Fulconis,
  Duligall, Wadsworth, and Russell}}]{Rarity05}
\bibinfo{author}{\bibfnamefont{J.~G.} \bibnamefont{Rarity}},
  \bibinfo{author}{\bibfnamefont{J.}~\bibnamefont{Fulconis}},
  \bibinfo{author}{\bibfnamefont{J.}~\bibnamefont{Duligall}},
  \bibinfo{author}{\bibfnamefont{W.~J.} \bibnamefont{Wadsworth}},
  \bibnamefont{and} \bibinfo{author}{\bibfnamefont{P.~S.~J.}
  \bibnamefont{Russell}}, \bibinfo{journal}{Opt. Express}
  \textbf{\bibinfo{volume}{13}}, \bibinfo{pages}{534} (\bibinfo{year}{2005}).

\bibitem[{\citenamefont{Hansryd et~al.}(2002)\citenamefont{Hansryd, Andrekson,
  Westlund, Li, and Hedekvist}}]{Hans02}
\bibinfo{author}{\bibfnamefont{J.}~\bibnamefont{Hansryd}},
  \bibinfo{author}{\bibfnamefont{P.~A.} \bibnamefont{Andrekson}},
  \bibinfo{author}{\bibfnamefont{M.}~\bibnamefont{Westlund}},
  \bibinfo{author}{\bibfnamefont{J.}~\bibnamefont{Li}}, \bibnamefont{and}
  \bibinfo{author}{\bibfnamefont{P.~O.} \bibnamefont{Hedekvist}},
  \bibinfo{journal}{IEEE J. Sel. Top. Quant.} \textbf{\bibinfo{volume}{8}},
  \bibinfo{pages}{506} (\bibinfo{year}{2002}).

\bibitem[{\citenamefont{Voss et~al.}(2006)\citenamefont{Voss,
  K\"{o}pr\"{u}l\"{u}, and Kumar}}]{Voss06}
\bibinfo{author}{\bibfnamefont{P.~L.} \bibnamefont{Voss}},
  \bibinfo{author}{\bibfnamefont{K.~G.} \bibnamefont{K\"{o}pr\"{u}l\"{u}}},
  \bibnamefont{and} \bibinfo{author}{\bibfnamefont{P.}~\bibnamefont{Kumar}},
  \bibinfo{journal}{J. Opt. Soc. Am. B} \textbf{\bibinfo{volume}{23}},
  \bibinfo{pages}{598} (\bibinfo{year}{2006}).

\bibitem[{\citenamefont{McKinstrie and Gordon}(2012)}]{Mck12}
\bibinfo{author}{\bibfnamefont{C.~J.} \bibnamefont{McKinstrie}}
  \bibnamefont{and} \bibinfo{author}{\bibfnamefont{J.~P.}
  \bibnamefont{Gordon}}, \bibinfo{journal}{IEEE J. Sel. Top. Quantum Electron.}
  \textbf{\bibinfo{volume}{18}}, \bibinfo{pages}{958} (\bibinfo{year}{2012}).

\bibitem[{\citenamefont{Gao et~al.}(1998)\citenamefont{Gao, Cui, Xue, Xie, and
  Peng}}]{Gao98}
\bibinfo{author}{\bibfnamefont{J.}~\bibnamefont{Gao}},
  \bibinfo{author}{\bibfnamefont{F.}~\bibnamefont{Cui}},
  \bibinfo{author}{\bibfnamefont{C.}~\bibnamefont{Xue}},
  \bibinfo{author}{\bibfnamefont{C.}~\bibnamefont{Xie}}, \bibnamefont{and}
  \bibinfo{author}{\bibfnamefont{K.}~\bibnamefont{Peng}},
  \bibinfo{journal}{Opt. Lett.} \textbf{\bibinfo{volume}{23}},
  \bibinfo{pages}{870} (\bibinfo{year}{1998}).

\bibitem[{\citenamefont{Su et~al.}(2006)\citenamefont{Su, Tan, Jia, Pan, Xie,
  and Peng}}]{Su06}
\bibinfo{author}{\bibfnamefont{X.}~\bibnamefont{Su}},
  \bibinfo{author}{\bibfnamefont{A.}~\bibnamefont{Tan}},
  \bibinfo{author}{\bibfnamefont{X.}~\bibnamefont{Jia}},
  \bibinfo{author}{\bibfnamefont{Q.}~\bibnamefont{Pan}},
  \bibinfo{author}{\bibfnamefont{C.}~\bibnamefont{Xie}}, \bibnamefont{and}
  \bibinfo{author}{\bibfnamefont{K.}~\bibnamefont{Peng}},
  \bibinfo{journal}{Opt. Lett.} \textbf{\bibinfo{volume}{31}},
  \bibinfo{pages}{1133} (\bibinfo{year}{2006}).

\bibitem[{\citenamefont{Funk and Raymer}(2002)}]{Funk02}
\bibinfo{author}{\bibfnamefont{A.~C.} \bibnamefont{Funk}} \bibnamefont{and}
  \bibinfo{author}{\bibfnamefont{M.~G.} \bibnamefont{Raymer}},
  \bibinfo{journal}{Phys. Rev. A} \textbf{\bibinfo{volume}{65}},
  \bibinfo{pages}{042307} (\bibinfo{year}{2002}).

\bibitem[{\citenamefont{Shelby et~al.}(1986)\citenamefont{Shelby, Levenson,
  Perlmutter, DeVoe, and Walls}}]{Shelby86}
\bibinfo{author}{\bibfnamefont{R.~M.} \bibnamefont{Shelby}},
  \bibinfo{author}{\bibfnamefont{M.~D.} \bibnamefont{Levenson}},
  \bibinfo{author}{\bibfnamefont{S.~H.} \bibnamefont{Perlmutter}},
  \bibinfo{author}{\bibfnamefont{R.~G.} \bibnamefont{DeVoe}}, \bibnamefont{and}
  \bibinfo{author}{\bibfnamefont{D.~F.} \bibnamefont{Walls}},
  \bibinfo{journal}{Phys. Rev. Lett.} \textbf{\bibinfo{volume}{57}},
  \bibinfo{pages}{691} (\bibinfo{year}{1986}).

\bibitem[{\citenamefont{Silberhorn et~al.}(2001)\citenamefont{Silberhorn, Lam,
  Wei\ss{}, K\"onig, Korolkova, and Leuchs}}]{Siberborn01}
\bibinfo{author}{\bibfnamefont{C.}~\bibnamefont{Silberhorn}},
  \bibinfo{author}{\bibfnamefont{P.~K.} \bibnamefont{Lam}},
  \bibinfo{author}{\bibfnamefont{O.}~\bibnamefont{Wei\ss{}}},
  \bibinfo{author}{\bibfnamefont{F.}~\bibnamefont{K\"onig}},
  \bibinfo{author}{\bibfnamefont{N.}~\bibnamefont{Korolkova}},
  \bibnamefont{and} \bibinfo{author}{\bibfnamefont{G.}~\bibnamefont{Leuchs}},
  \bibinfo{journal}{Phys. Rev. Lett.} \textbf{\bibinfo{volume}{86}},
  \bibinfo{pages}{4267} (\bibinfo{year}{2001}).

\bibitem[{\citenamefont{Hirosawa et~al.}(2005)\citenamefont{Hirosawa,
  Furumochi, Tada, Kannari, Takeoka, and Sasaki}}]{Hirosawa05}
\bibinfo{author}{\bibfnamefont{K.}~\bibnamefont{Hirosawa}},
  \bibinfo{author}{\bibfnamefont{H.}~\bibnamefont{Furumochi}},
  \bibinfo{author}{\bibfnamefont{A.}~\bibnamefont{Tada}},
  \bibinfo{author}{\bibfnamefont{F.}~\bibnamefont{Kannari}},
  \bibinfo{author}{\bibfnamefont{M.}~\bibnamefont{Takeoka}}, \bibnamefont{and}
  \bibinfo{author}{\bibfnamefont{M.}~\bibnamefont{Sasaki}},
  \bibinfo{journal}{Phys. Rev. Lett.} \textbf{\bibinfo{volume}{94}},
  \bibinfo{pages}{203601} (\bibinfo{year}{2005}).

\bibitem[{\citenamefont{Shelby et~al.}(1985)\citenamefont{Shelby, Levenson, and
  Bayer}}]{Shelby85}
\bibinfo{author}{\bibfnamefont{R.~M.} \bibnamefont{Shelby}},
  \bibinfo{author}{\bibfnamefont{M.~D.} \bibnamefont{Levenson}},
  \bibnamefont{and} \bibinfo{author}{\bibfnamefont{P.~W.} \bibnamefont{Bayer}},
  \bibinfo{journal}{Phys. Rev. B} \textbf{\bibinfo{volume}{31}},
  \bibinfo{pages}{5244} (\bibinfo{year}{1985}).

\bibitem[{\citenamefont{Guo et~al.}(in preparation)\citenamefont{Guo, Li, Liu,
  and Ou}}]{guo12}
\bibinfo{author}{\bibfnamefont{X.}~\bibnamefont{Guo}},
  \bibinfo{author}{\bibfnamefont{X.}~\bibnamefont{Li}},
  \bibinfo{author}{\bibfnamefont{N.}~\bibnamefont{Liu}}, \bibnamefont{and}
  \bibinfo{author}{\bibfnamefont{Z.~Y.} \bibnamefont{Ou}} (\bibinfo{year}{in
  preparation}).

\bibitem[{\citenamefont{Fabre et~al.}(1989)\citenamefont{Fabre, Giacobino,
  Heidmann, and Reynaud}}]{Fabre89}
\bibinfo{author}{\bibfnamefont{C.}~\bibnamefont{Fabre}},
  \bibinfo{author}{\bibfnamefont{E.}~\bibnamefont{Giacobino}},
  \bibinfo{author}{\bibfnamefont{A.}~\bibnamefont{Heidmann}}, \bibnamefont{and}
  \bibinfo{author}{\bibfnamefont{S.}~\bibnamefont{Reynaud}},
  \bibinfo{journal}{J. Phys. France} \textbf{\bibinfo{volume}{50}},
  \bibinfo{pages}{1209} (\bibinfo{year}{1989}).

\bibitem[{\citenamefont{Yuen and Chan}(1983)}]{Yuen83}
\bibinfo{author}{\bibfnamefont{H.~P.} \bibnamefont{Yuen}} \bibnamefont{and}
  \bibinfo{author}{\bibfnamefont{V.~W.~S.} \bibnamefont{Chan}},
  \bibinfo{journal}{Opt. Lett.} \textbf{\bibinfo{volume}{8}},
  \bibinfo{pages}{177} (\bibinfo{year}{1983}).

\bibitem[{\citenamefont{Inoue and Mukai}(2002)}]{Inoue02}
\bibinfo{author}{\bibfnamefont{K.}~\bibnamefont{Inoue}} \bibnamefont{and}
  \bibinfo{author}{\bibfnamefont{T.}~\bibnamefont{Mukai}}, \bibinfo{journal}{J.
  Lightwave Technol.} \textbf{\bibinfo{volume}{20}}, \bibinfo{pages}{969}
  (\bibinfo{year}{2002}).

\bibitem[{\citenamefont{Agrawal}(1995)}]{Agrawal}
\bibinfo{author}{\bibfnamefont{G.~P.} \bibnamefont{Agrawal}},
  \emph{\bibinfo{title}{Nonlinear fiber optics}} (\bibinfo{publisher}{Academic
  press}, \bibinfo{year}{1995}).

\bibitem[{\citenamefont{Hsieh et~al.}(2007)\citenamefont{Hsieh, Wong, Murdoch,
  Coen, Vanholsbeeck, Leonhardt, and Harvey}}]{Hsieh07}
\bibinfo{author}{\bibfnamefont{A.~S.~Y.} \bibnamefont{Hsieh}},
  \bibinfo{author}{\bibfnamefont{G.~K.~L.} \bibnamefont{Wong}},
  \bibinfo{author}{\bibfnamefont{S.~G.} \bibnamefont{Murdoch}},
  \bibinfo{author}{\bibfnamefont{S.}~\bibnamefont{Coen}},
  \bibinfo{author}{\bibfnamefont{F.}~\bibnamefont{Vanholsbeeck}},
  \bibinfo{author}{\bibfnamefont{R.}~\bibnamefont{Leonhardt}},
  \bibnamefont{and} \bibinfo{author}{\bibfnamefont{J.~D.}
  \bibnamefont{Harvey}}, \bibinfo{journal}{Opt. Express}
  \textbf{\bibinfo{volume}{15}}, \bibinfo{pages}{8104} (\bibinfo{year}{2007}).

\bibitem[{\citenamefont{Lali-Dastjerdi
  et~al.}(2012)\citenamefont{Lali-Dastjerdi, Rottwitt, Galili, and
  Peucheret}}]{Peucheret12}
\bibinfo{author}{\bibfnamefont{Z.}~\bibnamefont{Lali-Dastjerdi}},
  \bibinfo{author}{\bibfnamefont{K.}~\bibnamefont{Rottwitt}},
  \bibinfo{author}{\bibfnamefont{M.}~\bibnamefont{Galili}}, \bibnamefont{and}
  \bibinfo{author}{\bibfnamefont{C.}~\bibnamefont{Peucheret}},
  \bibinfo{journal}{Opt. Express} \textbf{\bibinfo{volume}{20}},
  \bibinfo{pages}{15530} (\bibinfo{year}{2012}).

\bibitem[{\citenamefont{Garay-Palmett et~al.}(2007)\citenamefont{Garay-Palmett,
  McGuinness, Cohen, Lundeen, Rangel-Rojo, U'ren, Raymer, McKinstrie, Radic,
  and Walmsley}}]{Palmett07}
\bibinfo{author}{\bibfnamefont{K.}~\bibnamefont{Garay-Palmett}},
  \bibinfo{author}{\bibfnamefont{H.~J.} \bibnamefont{McGuinness}},
  \bibinfo{author}{\bibfnamefont{O.}~\bibnamefont{Cohen}},
  \bibinfo{author}{\bibfnamefont{J.~S.} \bibnamefont{Lundeen}},
  \bibinfo{author}{\bibfnamefont{R.}~\bibnamefont{Rangel-Rojo}},
  \bibinfo{author}{\bibfnamefont{A.~B.} \bibnamefont{U'ren}},
  \bibinfo{author}{\bibfnamefont{M.~G.} \bibnamefont{Raymer}},
  \bibinfo{author}{\bibfnamefont{C.~J.} \bibnamefont{McKinstrie}},
  \bibinfo{author}{\bibfnamefont{S.}~\bibnamefont{Radic}}, \bibnamefont{and}
  \bibinfo{author}{\bibfnamefont{I.~A.} \bibnamefont{Walmsley}},
  \bibinfo{journal}{Opt. Express} \textbf{\bibinfo{volume}{15}},
  \bibinfo{pages}{14870} (\bibinfo{year}{2007}).

\end{thebibliography}

\end{document}